\documentstyle [aps,twocolumn,epsfig]{revtex}
\catcode`\@=11

\newcommand{\be}{\begin{equation}}
\newcommand{\ee}{\end{equation}}
\newcommand{\bea}{\begin{eqnarray}}
\newcommand{\eea}{\end{eqnarray}}

\newcommand{\s}{\sigma}
\newcommand{\rsi}{\rm \sigma}

\newcommand{\rc}{{\rm c}}

\newcommand{\rj}{{\rm j}}

\begin{document}

\title{A Pseudogap in the 
Single-Particle Density of States of a Tomonaga-Luttinger Liquid}
\author{Emiliano Papa and  Alexei M. Tsvelik \\
{\em{Department of Theoretical Physics, University of Oxford,
1 Keble Road, Oxford, OX1 3NP}}}

\address{\rm (Received: )}
\address{\mbox{ }}
\address{\vspace{4mm}\parbox{14cm}{\rm \mbox{ }\mbox{ }
We study a single-particle density of states (DOS) 
in  a model of Tomonaga-Luttinger
liquid with small (large) values of the Luttinger parameter $K_\rc < 0.17$,
($K_\rc > 5.82$),
in the charge sector. We explain that such  values of $K_\rc$ may be  
achieved by  electron-phonon
interactions without generating the spin gap. 
We suggest that electron-phonon 
 interactions can be partially incorporated into the
Tomonaga-Luttinger liquid scheme by introducing frequency dependent
$K_\rc(\omega)$. 
We demonstrate that when the low-frequency asymptotic value $K_\rc(0) <
0.17$ or $K_\rc(0) > 5.82$,  the single-particle DOS has a pseudogap
behavior in frequency where $\rho(\omega)/\omega$ vanishes 
at small frequencies.
 The DOS exhibits a peak the position of which scales as 
$[K(0) + K^{-1}(0)]\omega_0/2$ 
 where $\omega_0$ is the  characteristic phonon frequency. 
}}
\address{\mbox{ }}
\address{\vspace{2mm}\parbox{14cm}{\rm PACS No: 71.10.Pm}}
\maketitle

\makeatletter
\global\@specialpagefalse
\makeatother

The Tomonaga-Luttinger liquid theory predicts a power-law frequency dependence
of the single-particle density of states (DOS). In an apparent 
 contradiction with this
prediction many quasi-one-dimensional materials (see, for example,
\cite{blue}) exhibit a pseudo-gap type structure of DOS.  

In this brief report we argue that it is perfectly possible 
to describe 
the pseudo-gap behavior in the Tomonaga-Luttinger framework provided
one  treats  carefully the short-time part  of the electron
Green's function. 

\sloppy

Let us recall how DOS is calculated. The standard 
non-perturbative technique used to study
low-dimensional systems - the bosonization approach yields thermodynamic Green's functions in the 
Matsubara time representation $G(\tau)$. 
To extract DOS one has to Fourier transform this function to get
$G(i\omega_n)$ and then do the 
analytic continuation $i\omega_n \rightarrow \omega + i0$.

As we have said, in the limit  $T
\rightarrow 0$,  the single-electron Green's function in the
Tomonaga-Luttinger theory has a power-law asymptotics:
\begin{equation}
G(\tau) =  \frac{A}{\tau}\frac{1}{|\tau|^{\theta}} 
\label{as}
\end{equation}
where $A$ is a non-universal number and the exponent 
$\theta$ is determined by the interactions. For the
standard spin-1/2 SU(2)-invariant Tomonaga-Luttinger liquid the
exponent $\theta$ is related to the Luttinger parameter $K_\rc$  
\cite{many}: 
\begin{equation}
\theta = \frac{1}{4}(K_\rc^{-1} + K_\rc - 2)
\end{equation}

Assuming that $G(- \tau) = - G(\tau)$ we can write down  the Fourier
transform as  
\begin{equation}
G(i\omega_n) = 2i\int_0^{\infty} d\tau \sin(\omega_n\tau)G(\tau) \label{four}
\end{equation}
At small frequencies this integral is dominated by
the asymptotics (\ref{as}) when $\theta < 1$. Then we have 
\begin{equation}
G(i\omega_n) \approx 2iA\int_0^{\infty}
d\tau \frac{\sin(\omega_n\tau)}{\tau^{1
+ \theta}} \sim (\omega_n)^{\theta}
\end{equation} 
The analytic continuation is quite straightforward in this case and 
one gets 
\bea
& &\Im m G^{(R)}(\omega) =
\nonumber \\ [3mm]
& = & A(\omega)^{\theta}\frac{\sin[\pi(1 -
\theta/2)]\sin[\pi(\theta - 1)/2]\Gamma(2 - \theta)}{\theta(\theta -
1)}  \quad .
\label{DOS}
\eea

 When $\theta > 1$ the integral (\ref{four}) converges at large
Matsubara times. However, at $\theta < 2$ its second 
derivative with respect to $\omega_n$ is
still determined by the asymptotics at large $\tau$   such that we get 
\begin{equation}
G(i\omega_n) = -i\omega_ng(0) - 2(\omega_n)^{\theta}\frac{\sin[\pi(1 -
\theta/2)]\Gamma(2 - \theta)}{\theta(\theta - 1)} ... 
\label{conj}
\end{equation}
After the analytic continuation only the second term contributes to
the imaginary part and we find that Eq. (\ref{DOS}) is still
valid. Thus the value $\theta =1$ corresponding either to $K_\rc = 3 - 2\sqrt 2
\approx 0.17$ or $K_\rc = 3 + 2\sqrt 2 \approx  5.83$ marks a crossover
 into a pseudogap phase  where $\rho(\omega)/\omega$ vanishes at
$\omega \rightarrow 0$.  

  Small values of $K_\rc$ can be achieved in systems  with strong
 retardation effects such as systems with electron-phonon
 interactions or Kondo lattices \cite{Ueda}. In all these systems
 small values of $K_\rc$ are achieved asymptotically at small
 frequencies. Therefore it is interesting to learn what is the area of
 validity of the universal power law behavior (\ref{DOS}). Below we
 calculate the DOS for a model with electron-phonon interactions. 

The influence of phonons on the electron subsystem 
in quasi-one-dimensional metals  have 
been studied by different authors; mostly
for the case of noninteracting electrons (see \cite{Heeger},
\cite{Grunner} and references therein). 
The combined effects of the Coulomb and electron- phonon interactions 
 was studied in \cite{Voit1},\cite{Voit2}  using the 
renormalization group approach.  
In \cite{GNT} these effects  have been studied in the framework of 
Tomonaga-Luttinger theory.

 Here we briefly repeat the derivation given in \cite{GNT}. 
The lattice effects can be included in the Hubbard Hamiltonian by making
the hopping integral $t$ dependent on the intersite distance:
\[
t_{\rm ij}\approx t+\frac{1}{2a}\kappa (u_{\rm i}-u_{\rm j})
\quad .
\]
(It has been proposed for the first time by Su, Schriefer and Heeger
 to describe the essential physics of  conducting polymers
\cite{Su}).
Then  the Hamiltonian takes the form:
\begin{eqnarray}
& H & =  -t\sum_{\rm j,\sigma}\left(c_{\rm j+1,\sigma}^{+}c_{\rm j,\sigma}
+c_{\rm j,\sigma}^{+}c_{\rm j+1,\sigma}\right)
 +  U\sum_{\rm j }n_{\rm j \uparrow} n_{\rm j \downarrow}
- \nonumber \\ [3mm]
& - &\frac{1}{2a}\kappa \sum_{\rm j,\sigma}
\left(u_{\rm j}-u_{\rm j+1}\right)
\left(c_{\rm j+1,\sigma}^{+}c_{\rm j,\sigma}
+c_{\rm j,\sigma}^{+}c_{\rm j+1,\sigma}\right)
+H_{\rm ph} \ ,
\nonumber
\end{eqnarray}
where $u_{\rm j}$ is dimensionless and $\kappa$ has dimensions of
energy.
The $c_{j,\rsi}$ operators are the usual creation and annihilation
operators for the electrons with spin $\sigma$ in the Wannier orbitals
at site $j$ and $n_{\rj,\s}$ is the number
of electrons. U is the repulsion of two electrons on the same site.

In the case of an  incommensurate band filling  
($4 k_{\rm F}\not= 2\pi/a$)  in the continuous approximation the
electron-phonon part of this Hamiltonian  generates a coupling 
between the lattice
deformations and the $2 k_{\rm F}$ and $4 k_{\rm F}$ components of the charge
density. The electron-phonon interaction  contributes to an
effectively retarded interaction between the electronic
densities:
\begin{equation}
S_{\rm int}=-\int {\rm d} \tau {\rm d} \tau{'} {\rm d} x
\sum_{l=1,2}\rho(2lk_{\rm F},x)D_l(\tau-\tau')\rho(2lk_{\rm F},x)
\ ,
\end{equation}
where $D_l(\tau)$ is the phonons Green's function at $q = 2lk_{\rm F}$
\cite{GNT}. Thus the main contribution to the interaction comes from
phonons with large  frequency. 
This interaction  effects both the spin and 
the charge sector. The phonons give a positive contribution to the
current-current coupling constant in the spin sector (i.e. $g_{\rm
s}^{(0)}\rightarrow g_{\rm s}$).
In this sector the electron-phonon interaction competes with repulsive
forces responsible for $g_{\rm 0}$. We assume that 
 the renormalized coupling constant is still repulsive such that there
 is no  spin gap. 

In the charge sector the phonons influence the dynamics as well as the
scaling dimensions, renormalizing the charge velocity $v_{\rm c}$ and $K_{\rm c}$.
 In the limit of small
frequencies $|\omega|\ll \omega_{\rm l}$ their  renormalized  values  are:
\begin{equation}
K_{\rm c}=\left(\frac{m}{m^{*}}\right)^{1/2}K_{\rm c}^{0}
\quad,\qquad 
\Large {\tilde v}_{\rm c}=\left(\frac{m}{m^{*}}\right)^{1/2}v_{\rm c}^{0}
\quad .
\end{equation}
where $m^*$ is interpreted as renormalized electrons mass 
with $m$ being the  bare mass. 

Since we are interested not only in the long time asymptotics of the 
Green's function, but in its  intermediate time behavior, we shall
model these behavior adopting the  modified Gaussian model 
with the time-dependent Luttinger parameter:
\begin{equation}
S=\frac{1}{2} \sum_{\omega ,q}\Phi_{\rm c}(-\omega,-q)\left[\frac{1}{v_{\rm c}}
\omega^2 f(i \omega)+v_{\rm c} q^2\right] \Phi_{\rm c}(\omega,q)
\quad ,
\end{equation}
where the function $f(\omega)$ takes values between $f(0)=m^*/m \equiv
K^{-2}$ and 
$f(\infty)=1$. In this brief report  we suggest a semi-phenomenological 
form for the function $f(\omega)$:
\begin{equation}
f(\omega)=1+\frac{\omega_{\rm 0}^2}{\omega^2+\omega_{\rm 0}^2}(K^{-2}-1)
\quad ,
\end{equation}

The above model yields the following Matsubara time single-electron 
Green's function
\bea
G(\tau)& = & -\frac{1}{\tau}\exp\left\{-\int_{0}^{\epsilon_{\rm F}/\omega_0}
\frac{{\rm d}x}{x}\left[\left(\frac{x^2+1}{x^2+K^{-2}}\right)^{1/4} -
\right. \right. \nonumber \\ [3mm]
& - &
\left. \left.
\left(\frac{x^2+K^{-2}}{x^2+1}\right)^{1/4} \right]^2\sin^2\left(\omega_0 \tau
x/2\right) \right\}
\quad .
\eea
where $\epsilon_{\rm F}$ is the ultraviolet cut-off (recall that we consider
the Green's function at coinciding spatial points). 
Pictures of the single
particle density of states are represented on Figure 1. They are obtained
by analytic continuation of $G(i\omega_{\rm n})$ from the imaginary axis
to just above the real axis.

The figures clearly show the crossover from the Luttinger-liquid type
behavior with singular $d\rho/d\omega$ at $K >  0.17$ to the
pseudogap behavior at $K < 0.17$. Another remarkable feature of these
DOS is the peak at $\omega \approx K^{-1}\omega_0/2$ (see Fig. 2). Since the DOS
is invariant under $K \rightarrow K^{-1}$ this empirical formula can
be generalized as
\begin{equation}
\omega_{\rm peak} = \frac{1}{2}(K^{-1} + K)\omega_0 
\label{peak}
\quad .
\end{equation}
It  is interesting
that the peak always 
occurs at frequencies larger than the characteristic
phonon frequency
$\omega_0$. At small (large) values of $K$ this discrepancy can be
quite substantial. For example, a  peak in DOS 
has been observed in K$_{0.3}$MoO$_3$ at
$\omega \approx 300$ meV. The behavior of DOS at smaller frequencies
is almost linear in $\omega$ which suggests $K \approx 0.15$. Then 
Eq. (\ref{peak}) gives  a
reasonable estimate for the phonon frequency: $\omega_0 \approx  90$ meV.  

E.P. is grateful to Steve Allen and particularly to Dave Allen  for helpful
discussions.

\newpage
\onecolumn

\begin{figure}
\unitlength=1mm
\begin{picture}(170,125.4)
\put(5,3){\line(1,0){160}}
\put(5,125.4){\line(1,0){160}}
\put(5,43.8){\line(1,0){160}}
\put(5,84.6){\line(1,0){160}}
%
\put(5,3){\line(0,1){122.4}}
\put(165,3){\line(0,1){122.4}}
\put(85,3){\line(0,1){122.4}}
\put(15,85.6){\epsfig{file=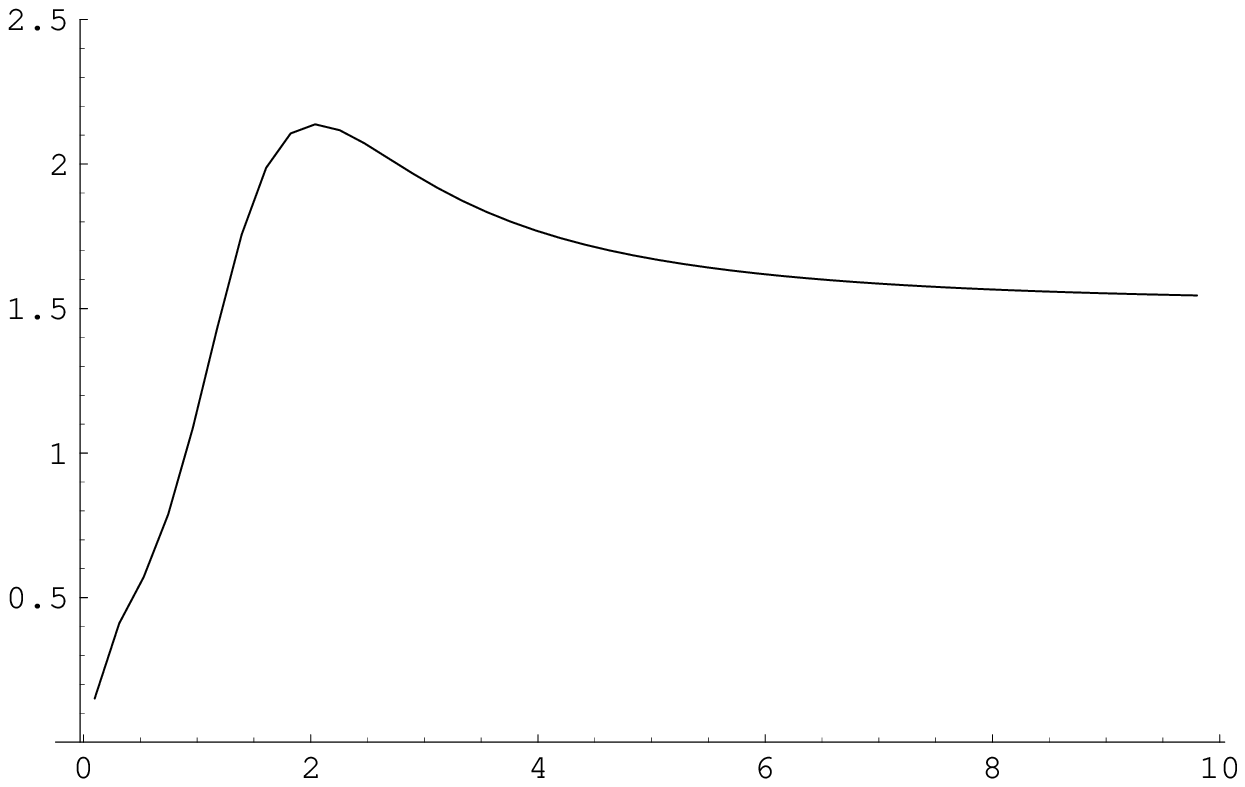,height=38mm}}
\put(95,85.6){\epsfig{file=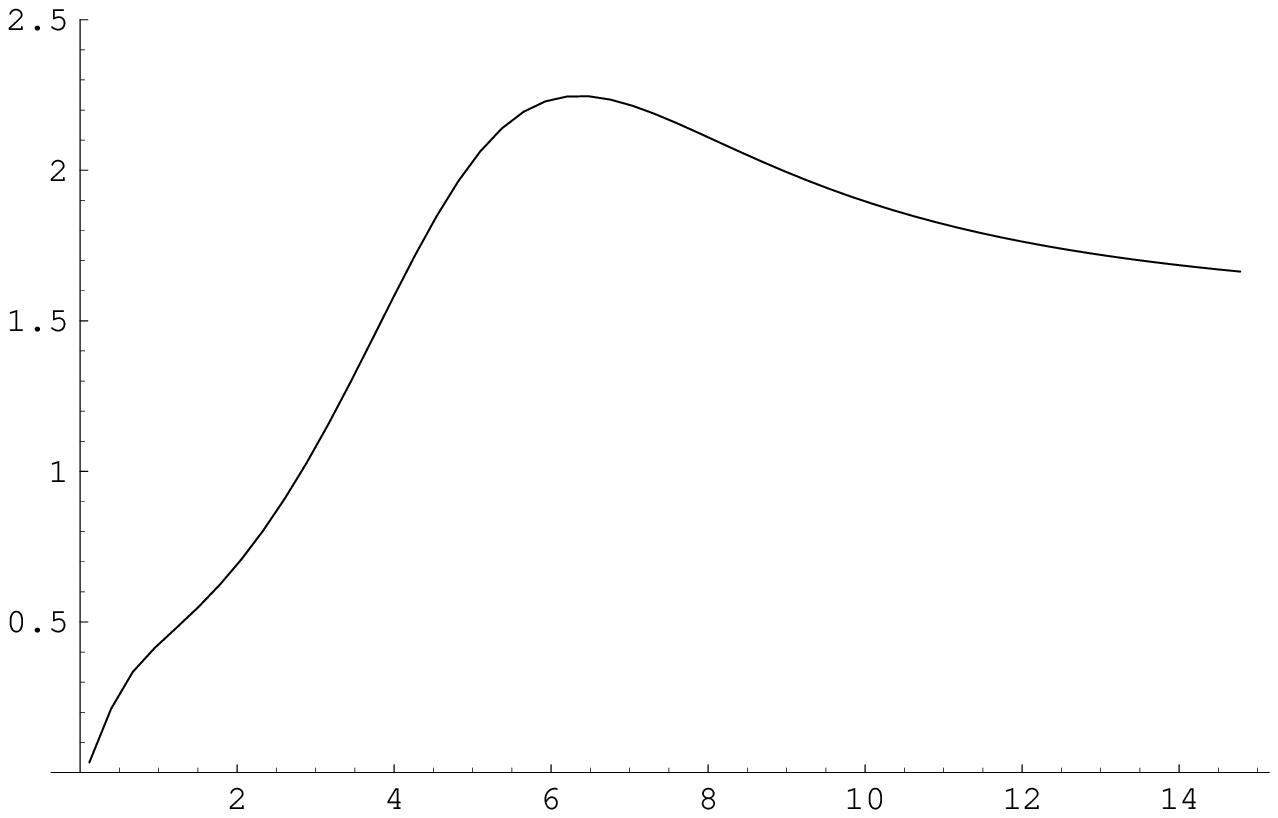,height=38mm}}
\put(15,44.8){\epsfig{file=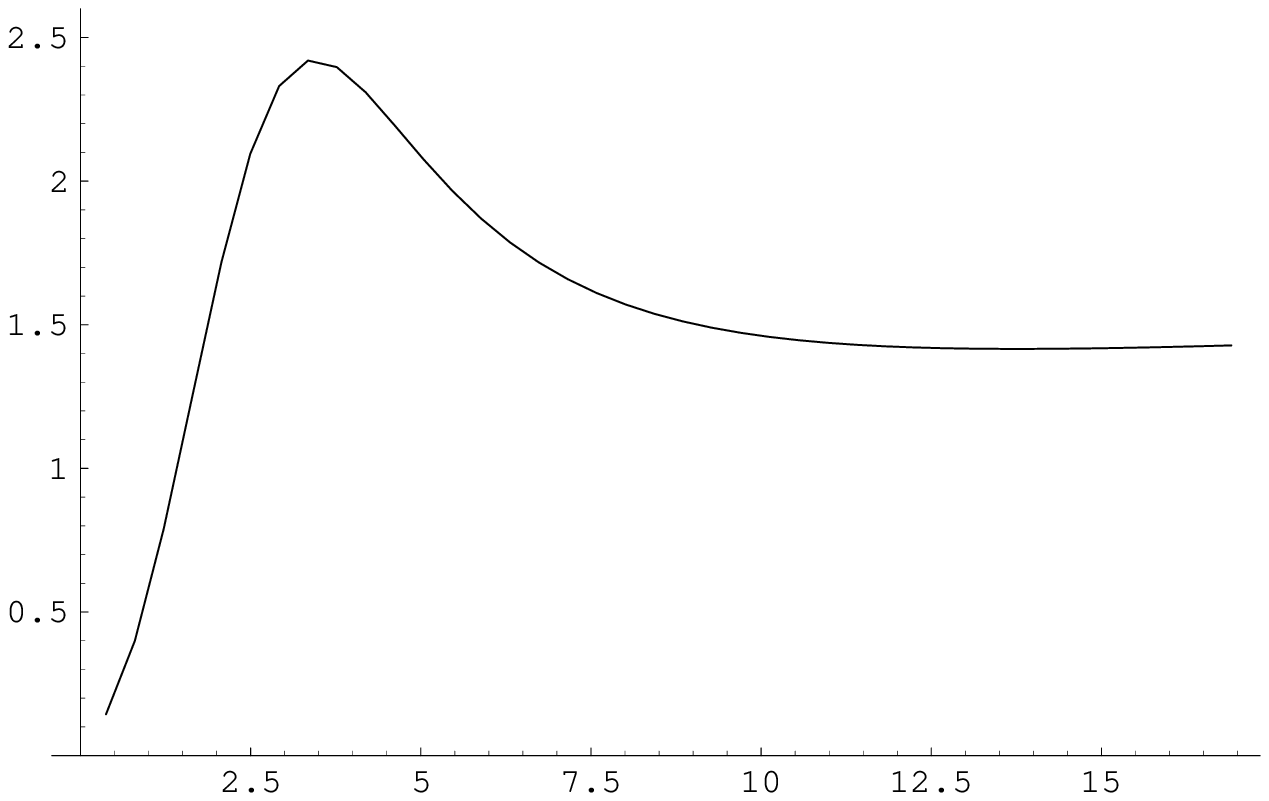,height=38mm}}
\put(95,44.8){\epsfig{file=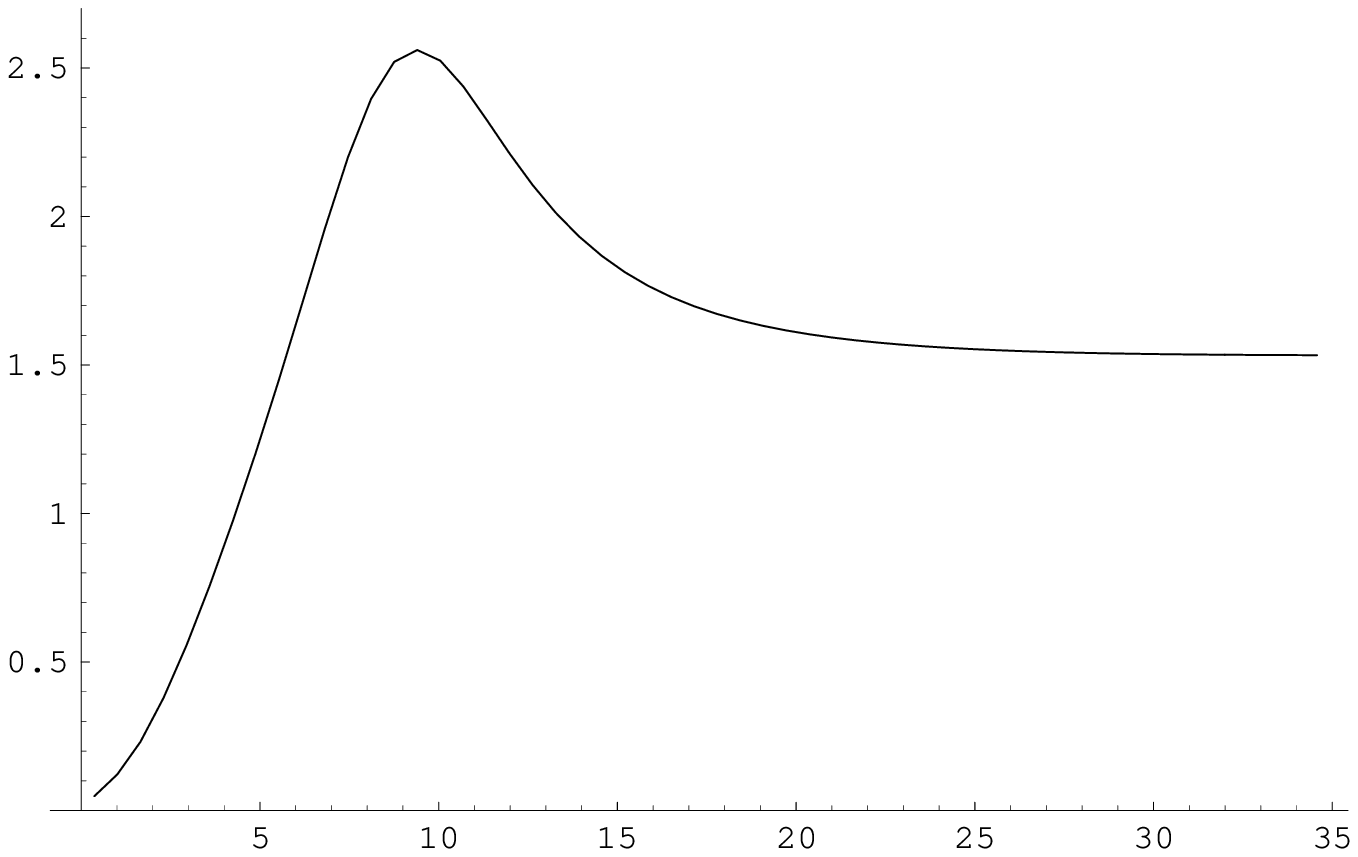,height=38mm}}
\put(15,4){\epsfig{file=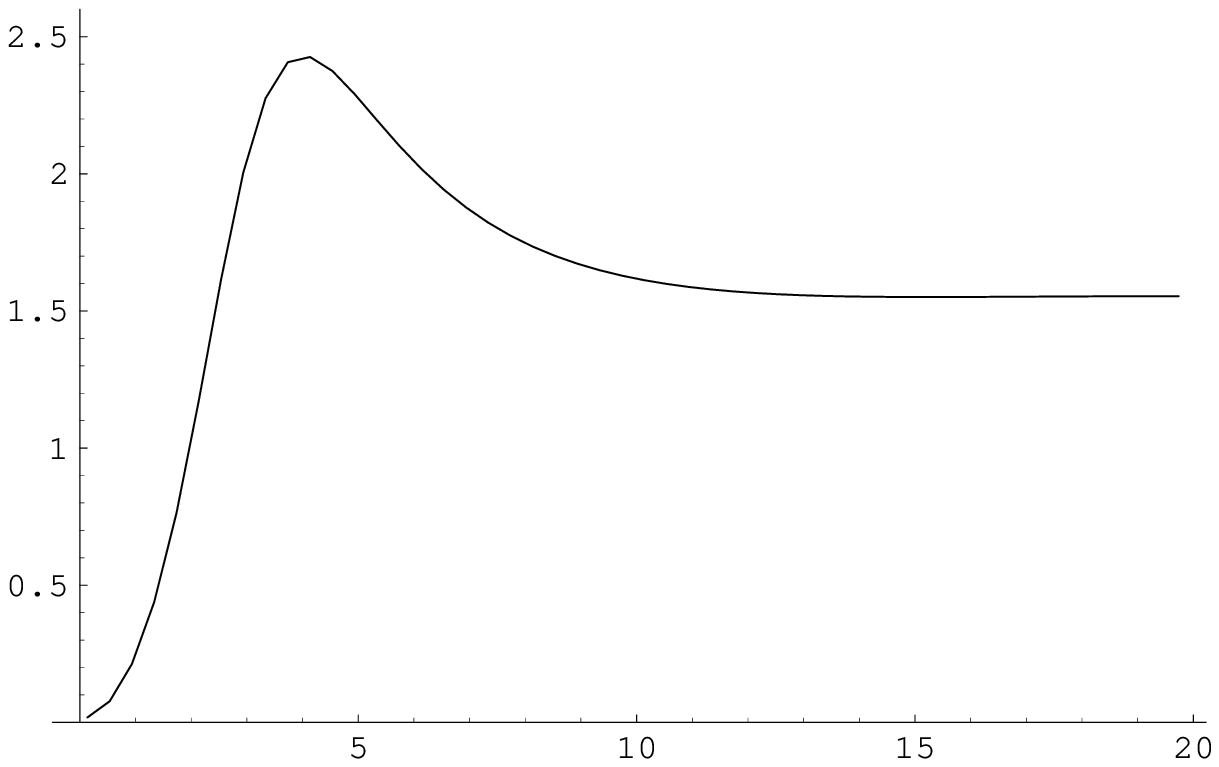,height=38mm}}
\put(95,4){\epsfig{file=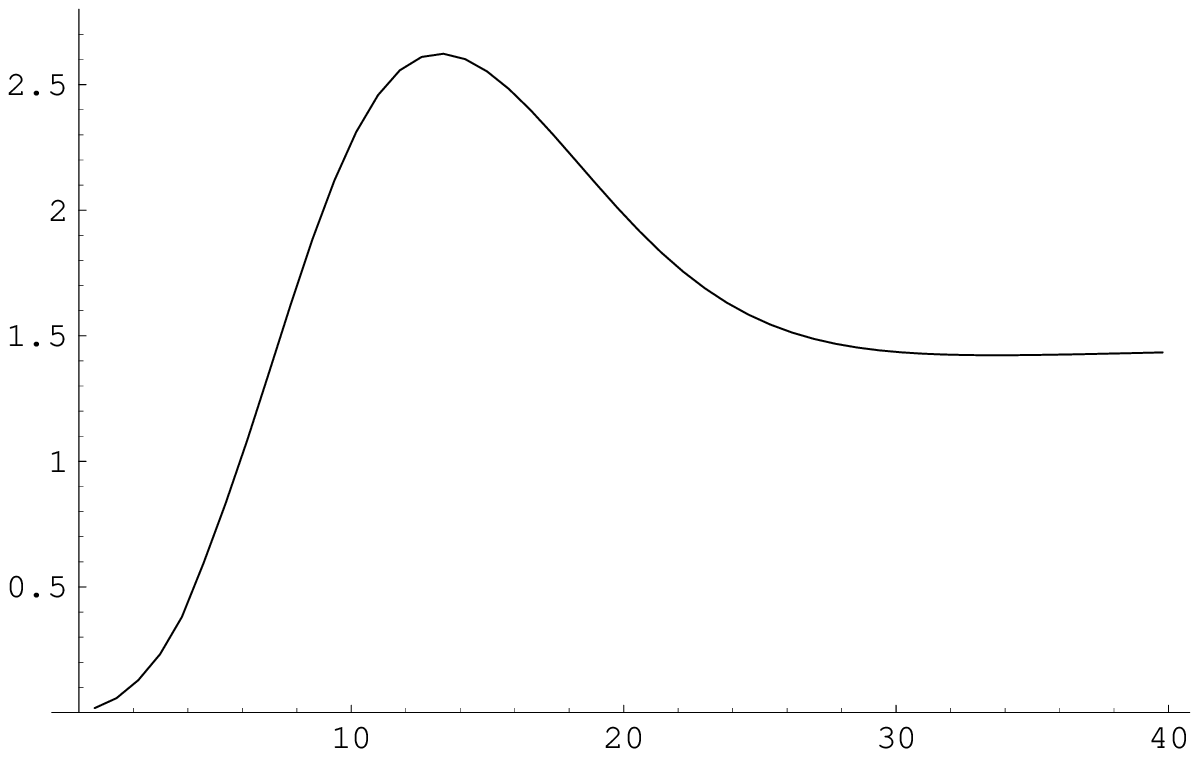,height=38mm}}
\put(12,30){\makebox(0,0)[cc]{$\rho(\omega)$}}
\put(12,70.8){\makebox(0,0)[cc]{$\rho(\omega)$}}
\put(12,111.6){\makebox(0,0)[cc]{$\rho(\omega)$}}
\put(92,30){\makebox(0,0)[cc]{$\rho(\omega)$}}
\put(92,70.8){\makebox(0,0)[cc]{$\rho(\omega)$}}
\put(92,111.6){\makebox(0,0)[cc]{$\rho(\omega)$}}
\put(80,7){\makebox(0,0)[cc]{$\omega$}}
\put(80,47.8){\makebox(0,0)[cc]{$\omega$}}
\put(80,88.6){\makebox(0,0)[cc]{$\omega$}}
\put(160,8){\makebox(0,0)[cc]{$\omega$}} 
\put(160,47.8){\makebox(0,0)[cc]{$\omega$}}
\put(160,88.6){\makebox(0,0)[cc]{$\omega$}}
\put(68.5,35){\makebox(0,0)[cc]{\underline{$K=1/9,\  \omega_0=1$}}}
\put(68.5,77.5){\makebox(0,0)[cc]{\underline{$K=1/7,\  \omega_0=1$}}}
\put(68.5,118.3){\makebox(0,0)[cc]{\underline{$K=1/5,\  \omega_0=1$}}}
\put(148.5,35){\makebox(0,0)[cc]{\underline{$K=1/9,\  \omega_0=3$}}}
\put(148.5,77.25){\makebox(0,0)[cc]{\underline{$K=1/7,\  \omega_0=3$}}}
\put(148.5,118.25){\makebox(0,0)[cc]{\underline{$K=1/5,\  \omega_0=3$}}}
\end{picture}
\caption{Plots of the dependence of the density of states on frequency
for different values of the parameters $K$ and $\omega_{\rm 0}$.}
\label{pseudogap}
\end{figure}

\begin{figure}
\unitlength=1mm
\begin{picture}(160,84.6)
\put(5,3){\line(1,0){160}}
\put(5,43.8){\line(1,0){160}}
\put(5,84.6){\line(1,0){160}}
%
\put(5,3){\line(0,1){81.6}}
\put(165,3){\line(0,1){81.6}}
\put(85,3){\line(0,1){81.6}}
\put(15,44.8){\epsfig{file=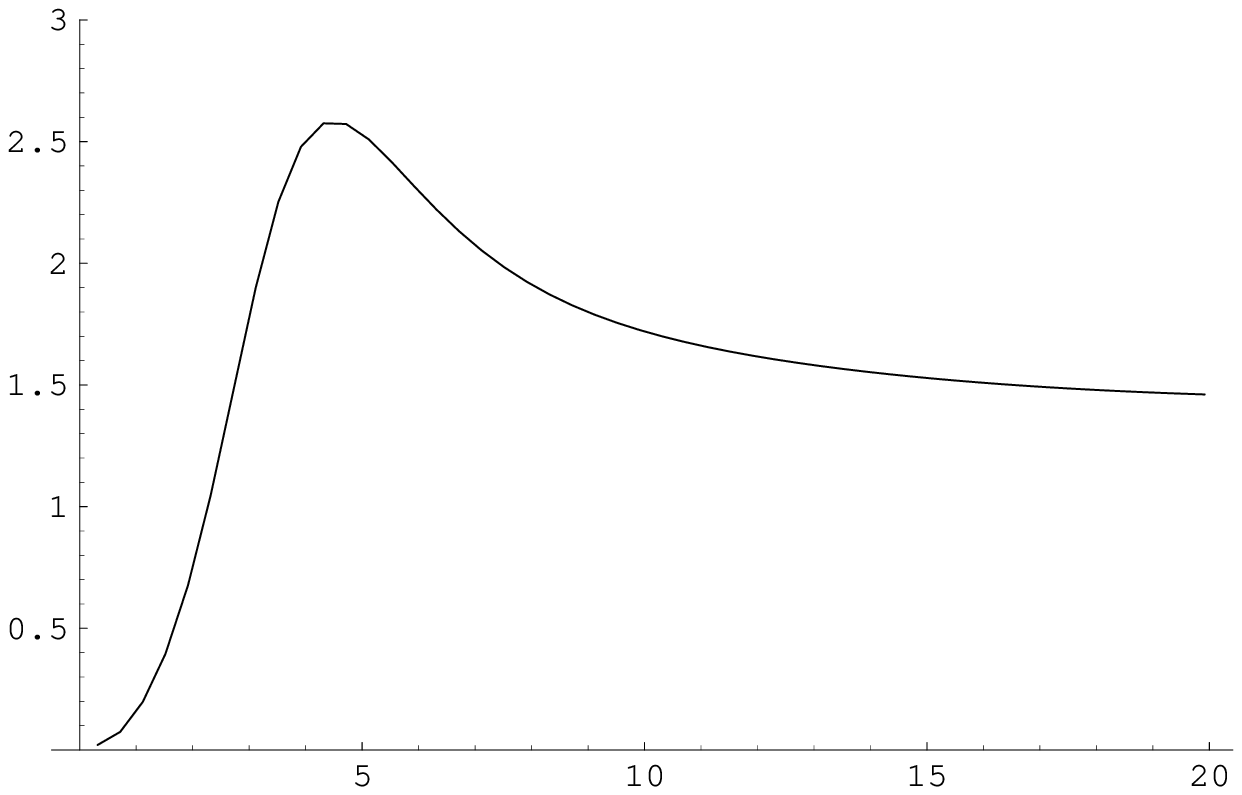,height=38mm}}
\put(95,44.8){\epsfig{file=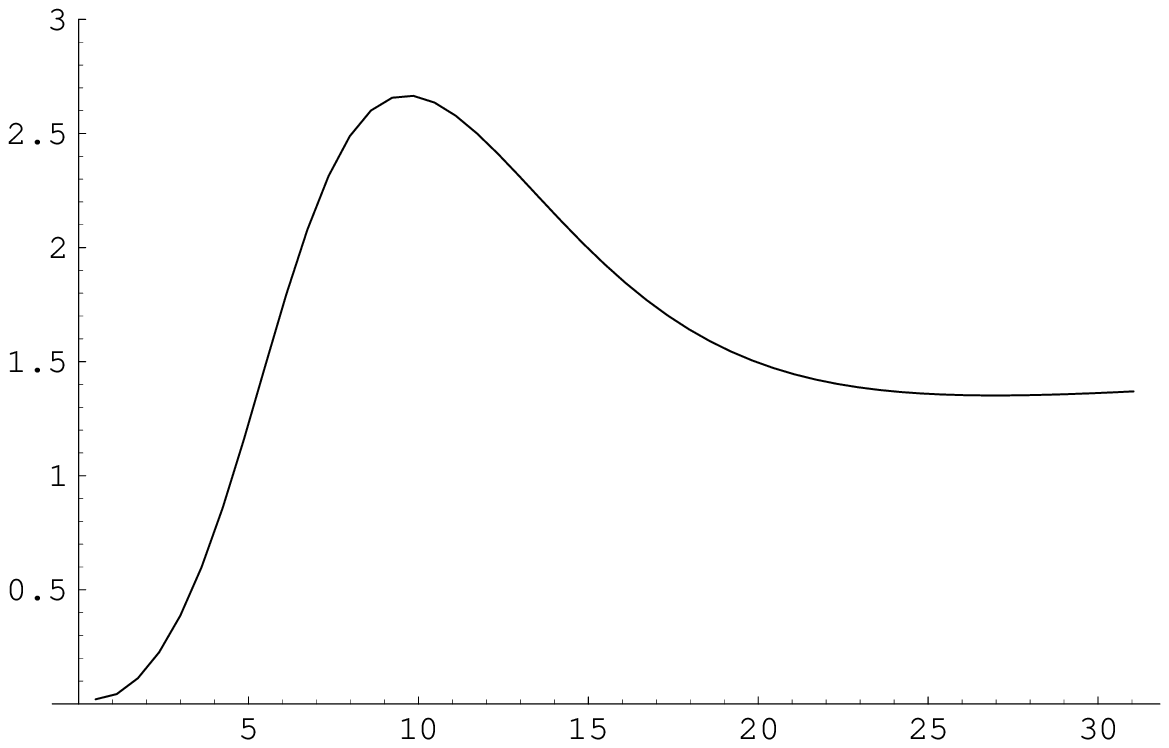,height=38mm}}
\put(15,4){\epsfig{file=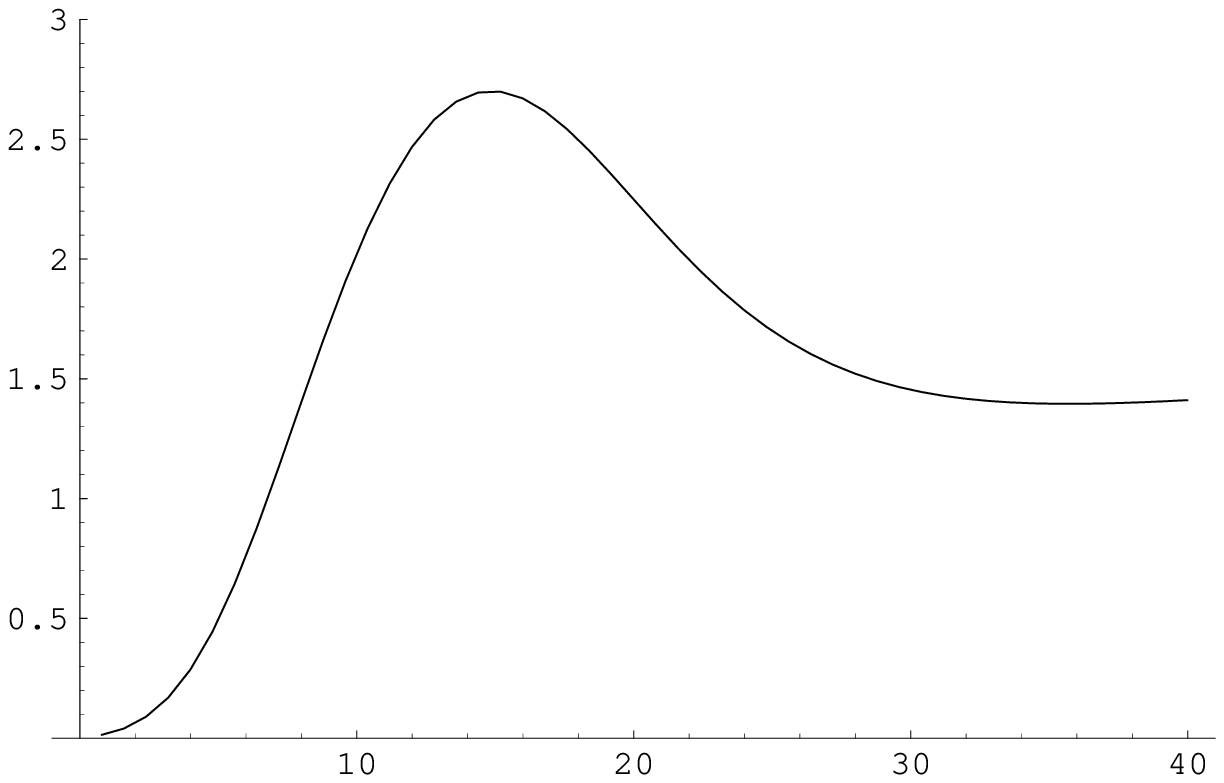,height=38mm}}
\put(95,4){\epsfig{file=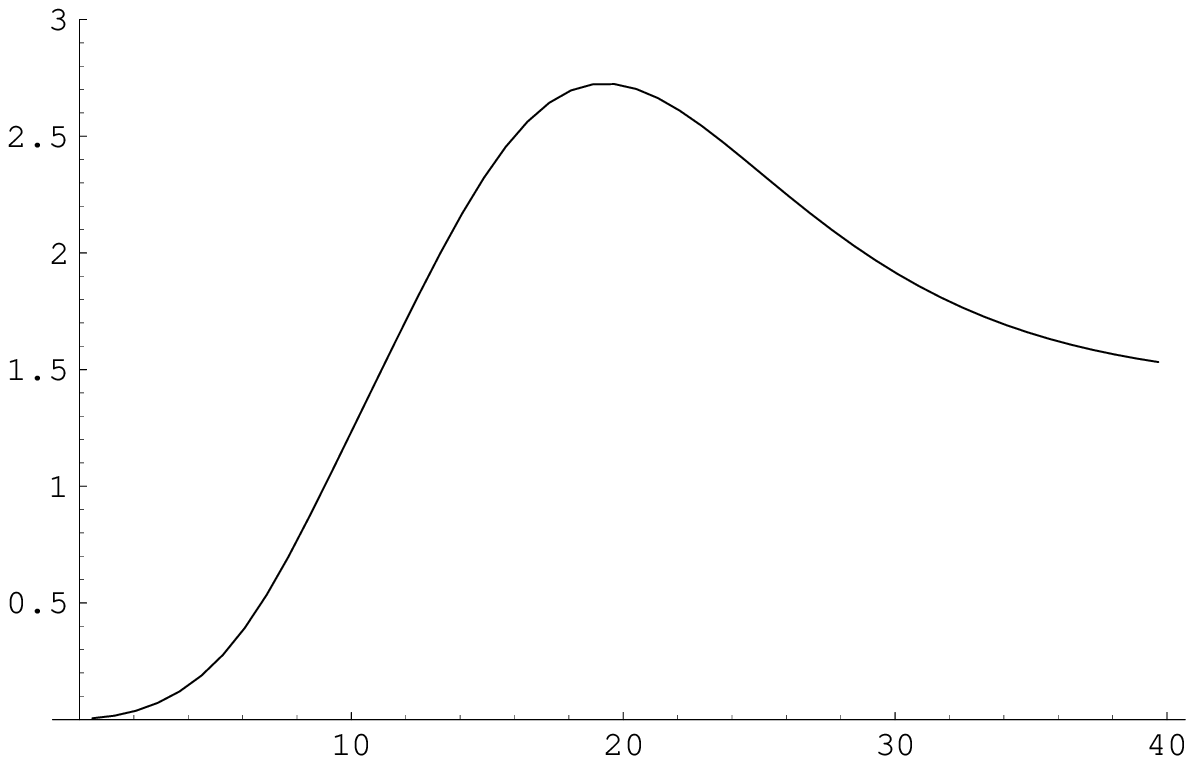,height=38mm}}
\put(12,30){\makebox(0,0)[cc]{$\rho(\omega)$}}
\put(12,70.8){\makebox(0,0)[cc]{$\rho(\omega)$}}
\put(92,30){\makebox(0,0)[cc]{$\rho(\omega)$}}
\put(92,70.8){\makebox(0,0)[cc]{$\rho(\omega)$}}
\put(80,7){\makebox(0,0)[cc]{$\omega$}}
\put(80,47.8){\makebox(0,0)[cc]{$\omega$}}
\put(160,7){\makebox(0,0)[cc]{$\omega$}} 
\put(160,47.8){\makebox(0,0)[cc]{$\omega$}}
\put(68.5,35){\makebox(0,0)[cc]{\underline{$K=0.1,\  \omega_0=3$}}}
\put(68.5,75.8){\makebox(0,0)[cc]{\underline{$K=0.1,\  \omega_0=1$}}}
\put(148.5,35){\makebox(0,0)[cc]{\underline{$K=0.1,\  \omega_0=4$}}}
\put(148.5,75.8){\makebox(0,0)[cc]{\underline{$K=0.1,\  \omega_0=2$}}}
\end{picture}
\caption{Plots of the dependence of the density of states on frequency
for different values of $\omega_0$ and fixed values of $K$. 
Here one can
observe the placing of the peak of $\rho(\omega)$ as function  of $\omega_0$.}
%
\label{maximum}
\end{figure}

\twocolumn

\end{document}